\documentclass[12pt]{article}
\usepackage{geometry}                		
\usepackage{graphicx}				
\usepackage{amsmath}								
\usepackage{amssymb}
\usepackage{amscd}
\usepackage{yfonts}
\usepackage{amsmath,amsthm,amssymb}
\usepackage{amsthm}
\newtheorem{theorem}{Theorem}

\newtheorem{definition}[theorem]{Definition}

              \newtheorem{lemma}[theorem]{Lemma}

\newcommand {\cH}{{\cal H}}
\newcommand{\bv}{{\bf v}}
\newcommand {\bp} {{\bf p}}
\newcommand {\bq} {{\bf q}}
\newcommand {{\bx}} {{\bf x}}
\newcommand {{\bk}} {{\bf k}}
\newcommand {\bP} {{\bf  P}}
\newcommand {\ba}{ {{\bf a}}}
\newcommand {\bC}{{ {\mathbb{C}}}}
\newcommand {\bR} {{\mathbb{R}}}
\newcommand  {\cA} {\mathcal{A}}

\newcommand  {\bra} {\langle}
\newcommand  {\ket }{\rangle}

\date{}

\medskip

\title{Scattering matrix and inclusive scattering matrix in algebraic quantum field theory}
\author{Albert Schwarz}
\date{}							

\begin{document}
\author {  A. Schwarz\\ Department of Mathematics\\ 
University of 
California \\ Davis, CA 95616, USA,\\ schwarz @math.ucdavis.edu}


\maketitle

{\it To Maxim Kontsevich with love and admiration}

\vskip .2 in
\abstract{We study the scattering of particles and quasiparticles in the framework of algebraic quantum field theory. The main novelty is the construction of inclusive scattering matrix related to inclusive cross-sections. The inclusive scattering matrix can be expressed in terms of generalized Green functions by a formula similar to the LSZ formula for the conventional scattering matrix.
The consideration of inclusive scattering matrix is necessary in quantum field theory if a unitary scattering matrix does not exist. It is always necessary if we want to consider the scattering of quasiparticles}
        \section {Introduction}
This paper was motivated by the wish to understand the scattering of quasiparticles (elementary
excitations of stationary translation-invariant state) in the framework of algebraic quantum field
theory (see \cite {S} for analysis of this problem in the framework of perturbation theory). It is
important to notice that our results are relevant also in quantum field theory, especially in
the case when the theory does not have particle interpretation ( the unitary scattering matrix
does not exist). The theory of scattering of relativistic particles (elementary excitations of the
ground state) in local quantum field theory (Haag-Ruelle theory) was developed in 1960th ( see \cite {AH} for the exposition closest to our approach).
 It was generalized to the non-relativistic
case in \cite {FS} (in this paper locality is replaced by asymptotic commutativity therefore the results
of it can be used also in string field theory \cite {STR}). Detailed exposition of the results of \cite { FS} was
given in the book \cite {MO}.
In the present paper we start with the generalization of some statements of the Haag-Ruelle
theory to the scattering of quasiparticles. The proofs are similar to the proofs in \cite {MO}. The
estimates that we are using can be obtained heuristically by stationary phase method, rigorous
proofs are either given in \cite {MO}or can be given by similar methods.
Although formally we can use Haag-Ruelle construction to define the scattering of quasiparticles this definition would not be reasonable. Quasiparticles are usually unstable; we should
consider only scattering of almost stable quasiparticles. However, it is very unlikely that at the end we have only almost stable quasiparticles. Therefore the conventional effective cross-section
is not well defined, we should consider the inclusive cross-section ( the probability that at the
end we have some almost stable quasiparticles plus something else.) We explain the way how
to calculate the inclusive cross-section of quasiparticle scattering. We introduce the notion
of generalized Green function in stationary translation-invariant state  $\omega$. ( These functions appear in the formalism of $L$-functionals \cite {SCH}, \cite {T},\cite {S}, \cite {MO}. They 
appear also in Keldysh formalism and in TFD (thermo-field dynamics); see, for example, \cite { UNI},
\cite {CU},\cite { K} for review of these formalisms.)
We define the matrix elements of inclusive scattering matrix as on-shell values of generalized
Green functions. We show that the inclusive cross-section of quasiparticles (of elementary
excitations of $\omega$) can be expressed in terms of matrix elements of the inclusive scattering
matrix. This statement can be useful also in the consideration of the scattering of particles
because quite often the experiment gives inclusive cross-section. It is necessary if a unitary
scattering matrix does not exist.
We are using the algebraic approach to quantum field theory. One can prove similar results
in the geometric approach to quantum theory suggested in \cite {GA1}, \cite {GA}.  Notice that in the geometric
approach it is necessary to consider the inclusive scattering matrix; it seems that the conventional scattering matrix cannot be defined in this approach.
In the appendix we discuss the scattering of unstable quasiparticles.
Notice that the definition of particle we are using is not always applicable; for example the
so called infraparticles (see \cite {BB}) are not particles in our sense. It seems that the right approach
in these cases is based on the techniques of \cite {SCH},\cite {S}.

The first version of present paper was published with small changes as Section 13.3 of \cite { MO}. The present version contains mostly  the same results, but in many places the exposition is 
different.  One of the goals of these changes is to alleviate the comparison of the results of this paper with more general results of \cite {GA2} and with the analysis of scattering in geometric approach  \cite {GA3} and  in  approach based on consideration  of Jordan algebras \cite {GA4}.
 \section{Algebraic approach to quantum theory}
 Quantum mechanics can be formulated in terms of an algebra of observables.
 The starting point of this formulation is a unital associative algebra $\cal A$ (the algebra of observables). In present section  this algebra can be an algebra over real or  complex numbers,  in  other  sections   we assume that $\cal A$ is an algebra over $\mathbb {C}$.
 
 One assumes that this algebra is equipped with antilinear involution $A\rightarrow A^*$.
 One says that a linear functional $\omega$ on $\cA$ specifies a state if $\omega (A^*A) \geq 0$ (i.e.\ if the functional is positive).
 The space of states  is a convex set that is invariant with respect to multiplication by a non-negative number  ( a convex cone). It will be denoted  by $\cal C$.  Proportional elements of $\cal C$ are identified, hence one can work with states obeying
 $\omega (1)=1 $  (with normalized states). 
 
 Sometimes we use the notation $\langle A\rangle_{\omega}$ or simply $\langle A\rangle$
 for $\omega (A)$; we say that this is the expectation value of $A$ in the state $\omega.$

 
 Sometimes we use the notation $\langle A\rangle_{\omega}$ or simply $\langle A\rangle$
 for $\omega (A)$; we say that this is the expectation value of $A$ in the state $\omega.$

 In the standard exposition of quantum mechanics, the algebra of observables
 consists of operators acting on a (pre-) Hilbert space.
 Every vector $x$ having a unit norm specifies a state by the formula
$\omega (A)= \bra Ax,x\ket$. Sometimes we are using bra-ket notations, in these notations $\omega (A)=\bra x| A|x \ket.$
 (More generally, a density matrix $K$ defines a state by the formula $\omega (A)={\rm Tr} AK$.)
 This situation is in some sense universal: for every state $\omega$ on $\cA$ one can construct a (pre-) Hilbert space $\cH$ and a representation of $\cA$ by operators on this space in such a way that the state $\omega$ corresponds to a vector in this space (GNS construction).

 To construct $\cH$ one defines inner product on $\cA$ by the formula $\bra A,B \ket=\omega (B^*A)$.
 The space $\cH$ can be obtained from $\cA$ by means of factorization with respect to zero vectors of this inner product.
 The inner product on $\cA$ descends to $\cH$, providing it with the structure of a pre-Hilbert space.
 The state $\omega$ is represented by a vector $\Phi$ of $\cH$ that corresponds to the unit element of $\cA$.
 An operator of multiplication from the left by an element $C\in \cA$ descends to an operator $\hat C$ acting on $\cH;$ this construction gives a representation of $\cA$ in the algebra of bounded operators on $\cH$.
 The algebra of operators of the form $\hat C$ where $C\in \cA$ will be denoted by $\hat {\cA}(\omega)$.
 (In our definition $\cH$ is a pre-Hilbert space; taking its completion we obtain a representation of $\cA$ by operators in Hilbert space $\bar {\cH}$.)

 Although every state of the algebra $\cA$ can be represented by a vector in Hilbert space, in general it is impossible to identify Hilbert spaces corresponding to different states.

 Time evolution in the algebraic formulation is specified by a one-parameter group $\alpha(t)$ of automorphisms of the algebra $\cA$ preserving the involution.
 This group acts in the obvious way on the space of states.
 If $\omega$ is a stationary state (a state invariant with respect to time evolution), then the group $\alpha(t)$ descends to a group $U(t)$ of unitary transformations of the corresponding space $\cH$.
 Representing  $U(t)$  as $e ^{-iHt}$ we can say that $ H$ plays the role of the Hamiltonian; it can be considered as a self-adjoint operator in $\bar{\cH}$ (or in complexification of this space if we are working over $\mathbb {R}$).
 The vector $\Phi$ representing $\omega$ obeys $H\Phi=0.$
 We say that the stationary state $\omega$ is a ground state if the spectrum of $H$ is non-negative.

 Every element $B$ of the algebra $\cA$ specifies two operators acting on linear functionals on $\cA$:
 \begin{equation}
  \label{eq:50.61}
  (B\omega)(A)= \omega (AB),
 \end{equation}
 \begin{equation}
  \label{eq:50.611}
  (\tilde B\omega)(A)=\omega (B^*A)
 \end{equation}

 Notice that the first of these operators is denoted by the same symbol as the element.
 The operator $B$ commutes with the operator $\tilde B'$.
 These operators do not preserve positivity, but the operator $\tilde B B$ does.
 This means that the operator sending $\omega$ into $\tilde B B(\omega)$ acts on the cone of states $\cal C$.
 If a state $\omega$ corresponds to a vector $\Phi$ in a representation of $\cA$ (i.e. $\omega(A)=\bra A\Phi,\Phi\ket$), then the state corresponding to the vector $B\Phi$ is equal to $\tilde B B \omega.$

 Symmetries of quantum theory in the algebraic formulation are automorphisms of the algebra $\cal A$ commuting with the involution and the evolution automorphisms $\alpha (t).$
 We are especially interested in the case when among the symmetries are
 operators $\alpha (\bx,t)$, where $\bx \in \bR ^d$ and $t\in \bR$, that are automorphisms of $\cA$ obeying $\alpha (\bx,t) \alpha (\bx ^{\prime},t ^{\prime}) = \alpha (\bx +\bx ^{\prime}, t+t ^{\prime})$. These automorphisms can be interpreted as space-time translations.
 
 We will use the notation $A(\bx,t)$ for $\alpha (\bx, t) A$, where $A \in \cA$.

One can say that the algebra $\cA$ and space-time translations specify a quantum theory in $\bR^d.$
 
 We will define particles as elementary excitations of the ground state; the theory of  particles and their collisions should be considered as quantum field theory  in  algebraic approach. (See the definition of elementary excitation   for the case when $\cA$ is an algebra over  $\bC$ in Section 3.)


 The action of a translation group on $\cA$ induces an action of this group on the space of states.
 In relativistic quantum field theory one should have an action of the Poincar\'e group on the algebra of observables (hence on states).


 Let us now consider a state $\omega$ that is invariant with respect to the translation group.
 
 We will define (quasi) particles as "elementary excitations" of $\omega$.
 To consider collisions of (quasi) particles, we should require that $\omega$ satisfies a cluster property in some sense.

 The weakest form of cluster property is the following condition
 \begin{equation}
  \label{eq:50.7}
  \omega (A(\bx, t)B)=\omega(A)\omega(B)+\rho(\bx,t)
 \end{equation}
 where $A,B\in \cA$ and $\rho$ is small in some sense for $\bx \to \infty$.
 For example, we can impose the condition that $\int |\rho (\bx, t)|d\bx< c(t)$, where $c(t)$ has at most polynomial growth.
 Notice that~\eqref{eq:50.7} implies asymptotic commutativity in some sense: $\omega ([A(\bx, t),B])$ is small for $\bx\to\infty.$

 To formulate a more general cluster property, we introduce the notion of correlation functions in the state $\omega:$
 \[w_n(\bx_1, t_1,\dots \bx_n,t_n)= \omega (A_1(\bx_1, t_1)\cdots A_n(\bx_n,t_n))=\langle  A_1(\bx_1, t_1)\cdots A_n(\bx_n,t_n)\rangle ,\]
 where $A_i\in\cA.$
 These functions generalize Wightman functions of relativistic quantum field theory.
 We consider the corresponding truncated correlation functions $w_n^T(\bx_1, t_1,\dots \bx_n,t_n)$ denoted also by $ \langle A_1(\bx_1, t_1)\cdots A_n(\bx_n,t_n)\rangle ^T.$
 
 Recall  that truncated correlation functions $w^T$ can be defined recursively by the formula
\begin{equation}
 w_n({\bx}_1,\tau_1,k_1\dots,{\bx}_n,\tau_n,k_n) = \sum_{s=1}^n \sum_{\rho \in R_s} w_{\alpha_1}^T(\pi_1) \dots w_{\alpha_s}^T(\pi_s). \label{eq:36.11}
\end{equation}
Here  $R_s$ denotes the collection of all  partitions  of the set $\{1,...,n\}$ into $s$ subsets denoted $\pi_1,...,\pi_s$, the number of elements in the subset $\pi_i$  is denoted by $\alpha_i$ and $w_{\alpha_i}^T(\pi_i)$ stands for  the truncated  correlation function   with arguments ${\bx}_a,\tau_a,k_a$ where $a\in \pi_i.$

 We have assumed that the state $\omega$ is translation-invariant; it follows that both correlation functions and truncated correlation functions depend on the differences $\bx_i-\bx_j, t_i-t_j.$
 We say that the state $\omega$ has the cluster property if the truncated correlation functions are small for $\bx_i-\bx_j\to \infty.$
 A strong version of the cluster property is the assumption that the truncated correlation functions tend to zero faster than any power of $\min \|\bx_i-\bx_j\|$.
 Then its Fourier transform with respect to variables $\bx_i$ has the form $\nu_n (\bp_2,\dots ,\bp_n, t_1,\dots ,t_n)\delta (\bp_1+\dots +\bp_n)$, where  $\nu_n$ is a smooth function of $\bp_2,...\bp_n$.

 Instead of cluster property, one can impose a condition of asymptotic commutativity of the algebra $\hat {\cal A}(\omega).$
 In other words, one should require that the commutator $[\hat A(\bx,t), \hat B]$ where $A,B\in\cA$ is small for $\bx\to \infty.$
 For example, we can assume that for every $n$ the norm of this commutator is bounded from above by $C_n(t) (1+\|x\|)^{-n}$, where the function $C_n(t)$ has at most polynomial growth.
 (This property will be called strong asymptotic commutativity.)
 Notice that in relativistic quantum field theory, in the Haag-Araki or the Wightman formulation, strong asymptotic commutativity can be derived from locality and the existence of mass gap.
 
  We say that a state $\sigma$ is an excitation of $\omega$ if it coincides with $\omega$ at infinity up to a constant factor.
 More precisely we should require that $\sigma (A(\bx, t))\to C\omega (A)$ as $\bx\to \infty$ for every $A\in \cal A$.
 Notice that the state corresponding to any vector $\hat A\Phi$, where $A\in \cal A$ is an excitation of $\omega$; this follows from the cluster property.

  \section{Particles and quasiparticles}

 In Sections 3 and 4 we show how one can define  elementary excitations of translation-invariant  state $\omega$   and the scattering of  elementary excitations ((quasi) particles)  for such algebras. In Section 5 we express scattering matrix and inclusive scattering matrix in terms of Green functions.

 The action of the translation group on $\cA$ generates a unitary
 representation of this group on the pre-Hilbert space $\cH$ constructed from $\omega$. We represent $T_{\bx}$ as $e^{i\bP\bx}$ and $T_{\tau}$ as $e^{-iH\tau}.$
 Operators  $\bP$ and $H$ are identified with the momentum operator and the Hamiltonian.
 
 The vector in the space $\cH$ that corresponds to $\omega$ will be denoted by $\Phi$.
 If $\Phi$ is a ground state (= the spectrum of  the Hamiltonian is non-negative) , we say that it is the physical vacuum.
 If $\omega$ obeys the KMS-condition $\omega ( A(t)B)=\omega (BA(t+i\beta))$,
 we say that $\omega$ is an equilibrium state with the temperature $T=\frac{1}{\beta}.$

 One can define a {\it one-particle state } (a one-particle excitation of the ground state $\omega$ or elementary excitation of $\omega$) as a generalized $\cH$-valued function $\Phi (\bp)$ obeying $\bP \Phi (\bp)=\bp \Phi (\bp),
H \Phi (\bp)=\varepsilon (\bp) \Phi (\bp)$.
 (More precisely, for some class of test functions $f(\bp)$ we should have a linear map $f\rightarrow \Phi (f)$ of this class into $\cH$ obeying $\bP \Phi (f)= \Phi (\bp f), H\Phi (f)=\Phi (\varepsilon (\bp) f)$, where $\varepsilon (\bp)$ is a real-valued function called dispersion law.
 For definiteness we  assume that test functions belong to the Schwartz space ${\cal {S}}(\bR ^d)$, i.e. to the space of smooth functions decreasing faster than any power.)
 
  We assume that $\Phi(f)$ is normalized (i.e. $\bra \Phi(f), \Phi(f')\ket=\bra f,f' \ket$). It follows that  one-particle state specifies an isometric  map ${\cal S}\to \cal H$ commuting with space and time translations. ( The spatial translation $T_{\bx}$ acts on $\cal S$ as multiplication by $e^{i\bp\bx}$ and the time translation $T_{\tau}$  acts as multiplication by $e^{-i\varepsilon (\bp)\tau}$.)
  
 {\it Let us fix an element $B\in \cal A$ such that} $\hat B\Phi=\Phi (\phi)$.
 (We assume that $\phi$ is smooth and does not vanish anywhere.)
 Recall that we consider $\cH$ as a pre-Hilbert space obtained by the GNS construction, therefore every element of $\cH$ can be represented in the form $\hat B\Phi$.

 Notice that for every function $g (\bx,t)\in {\cal {S} }(\bR ^{d+1})$ we have
 \begin{equation}
  \label{eq:50.8}
  \hat B(g)\Phi=\Phi(g_{\phi})
 \end{equation}
 where $\hat B(g)=\int g (\bx,t)\hat B (\bx,t)d\bx dt$, $g_f(\bp)=\hat g (\bp,-\varepsilon(\bp))\phi(\bp)$, and $\hat g$ stands for the Fourier transform of $g.$

 It follows that we  always assume that $\hat B=\int \alpha (\bx,t) \hat A(\bx,t)d\bx dt$, where $\alpha (\bx,t)\in {\cal S}(\bR^{d+1}), A\in \cA.$
 We  say that an operator $\hat B$ satisfying this assumption is a smooth operator. If a smooth operator transforms  $\Phi $ into one-particle state we say that it is a  good operator.
 For a smooth operator $\hat B(\bx,t)=\int \alpha (\bx'-\bx,t'-t)\hat A(\bx',t')d\bx' dt'$, hence this expression is a smooth function of $\bx$ and $t.$


 Notice that it is possible that there exist several types of particles $\Phi_r(f)$ with different dispersion laws.
 We say that the space spanned by $\Phi_r(f)$ is one-particle space and denote it $\cH_1$.
 If the theory is invariant with respect to spatial rotations then the infinitesimal rotations play the role of components of angular momentum.
 If $d=3$, a particle with spin $s$ can be described as a collection of $2s+1$ functions $\Phi_r (\bp)$ obeying $\bP \Phi _r(\bp)=\bp \Phi _r(\bp),
H \Phi _r(\bp)=\varepsilon (\bp) \Phi _r(\bp)$.
 The group of spatial rotations acts in the space spanned by these functions; this action is a tensor product of the standard action of rotations of the argument and irreducible $(2s+1)$-dimensional representation.
 (Here $s$ is half-integer, the representation is two-valued if $s$ is not an integer.)
 In relativistic theory, an irreducible subrepresentation of the representation of the Poincar\'e group in $\cH$ specifies a particle with spin.

Notice that the definition of elementary excitation can be applied also in the case when $\omega$ is an arbitrary stationary translation-invariant state. However, in the general case we should use the word  {\it quasi-particle} instead of the word  {\it particle}.  More precisely we defined stable particles and quasi-particles. We say that we are dealing with unstable particles and quasi-particles if  the equality $H \Phi (\bp)=\varepsilon (\bp) \Phi (\bp)$ is satisfied only approximately. Quasi-particles are in general unstable.
 \section{Scattering}
 \subsection {In- and out-states}
  Let us assume that we have several  types of (quasi) particles defined as generalized functions $\Phi_k(\bp)$ obeying $\bP\Phi_k=\bp\Phi_k(\bp), H\Phi_k(\bp)=\varepsilon_k(\bp) \Phi_k(\bp)$, where the functions $\varepsilon_k(\bp)$ are smooth and strictly convex. ( Convexity of these functions is irrelevant for most of our results. The requirement that functions are smooth can be relaxed. To guarantee that time translations act in the space $\cal S$ we should assume that these functions have at most polynomial growth.)
 
 Take some good operators $B_k\in \cA$, obeying $\hat B_k\Phi=\Phi _k(\phi_k).$
 Define $\hat B_k(f,t)$, where $f$ is a function of $\bp$ as $\int \tilde f(\bx,t)\hat B_k(\bx,t)d\bx$ and $\tilde f(\bx,t)$ is a Fourier transform of $f(\bp)e^{-i\varepsilon_k(\bp)t}$ with respect to $\bp.$ 
 
 Notice that
 \begin {equation}\label {ON}
 \hat B_k(f,t)\Phi=\Phi _k(f\phi_k)
 \end {equation}
 does not depend on $t.$
 
 Sometimes it is convenient to represent   $\hat B_k(f,t)$ in the form
 \begin {equation} \label {GOO}
 \hat B_k(f,t)= \int d\bp f(\bp)e^{-i\varepsilon_k(\bp)t} \hat B_k(\bp,t)
 \end {equation}
 where $  \hat B_k(\bp,t)=\int d\bx e^{-i\bp\bx}\hat B_k(\bx,t).$ (We understand $  \hat B_k(\bp,t)$ as a a generalized function of $\bp$.)
 
 Let us consider the vectors 
 \begin{equation}
  \label{eq:50.9}
  \Psi(k_1,f_1,\dots ,k_n,f_n|t_1,\dots ,t_n)=\hat B_{k_1}(f_1,t_1) \cdots \hat B_{k_n}(f_n,t_n)\Phi
 \end{equation}
 We assume that $f_1,\dots ,f_n$ have compact support.
 Let us introduce the notation $\bv_i(\bp)=\nabla \varepsilon _{k_i}(\bp)$.
 The set of all $\bv_i(\bp)$ such that $f_i(\bp)\neq 0$ will be denoted $U_i.$  If the sets $\overline {U_i}$ (the closures of $U_i$) do not overlap
 we say that $(f_1, ...,f_n)\in {\cal S}^n$ obeys $NO$ condition.  {\it We assume that the elements of ${\cal S}^n$ obeying $NO$ condition are dense in ${\cal S}^n$ (or at least that they span a dense subset)} This assumption is not restrictive. For example, if  there exists only one type of (quasi)particles it is satisfied if the dispersion law $\varepsilon(\bp)$ is strictly convex  on dense convex subset.

 If the $NO$ condition is satisfied then assuming the strong cluster property or asymptotic commutativity (precise formulation will be given below) {\it one can prove that the vector}~\eqref{eq:50.9} {\it has a limit as } $t_i\to \infty$ {\it or} $t_i\to -\infty$. These limits are called out-states (in-states) and denoted $ \Psi(k_1,f_1,\dots ,k_n,f_n|\pm \infty.)$
 The sets spanned by these limits will be denoted ${\cal D}_+$ or ${\cal D}_-.$

 Notice that the assumption that the sets $\overline {U_i}$ do not intersect ($NO$ condition) can be omitted if the space-time dimension is $\geq 4.$
 In these dimensions we can drop the $NO$ condition defining the sets $\cal D_{\pm}.$

 The existence of the limit of the vectors~\eqref{eq:50.9} allows us to define M\o ller matrices.
 We introduce the Hilbert space $\cH_{as}$ as a Fock representation for the operators $a^+_k(f),a_k(f)$ obeying canonical commutation relations. ( Let us emphasize that in our notations the operator $a_k(f)$ is linear with respect to $f$  and the operator  $a^+_k(f)$ is is conjugate to $a_k(\bar f)$, hence it is  also linear with respect to $f$.)

 We define M\o ller matrices $S_-$ and $S_+$ as linear operators defined on $\cH_{as}$ and taking values in $\bar {\cH}$ by the formula
 \begin{equation}\label{eq:50.10}
  \Psi (k_1,f_1,\dots ,k_n,f_n | \pm \infty) = S_{\pm}(a^+_{k_1}( f_1\phi_{k_1}) \dots a^+_{k_n}( f_n \phi_{k_n})\theta)
 \end{equation}
 where $\theta$ stands for Fock vacuum.
 
  This formula specifies $S_{\pm}$ on a dense subset of the Hilbert space $\cH_{as}$.
 These operators are isometric (this will be derived from cluster property), hence they can be extended to $\cH_{as}$ by continuity.

 One can say that the vector
 \begin{equation}
  \label{eq:50.101}
  e^{-iHt} \Psi (k_1,f_1,\dots ,k_n,f_n|\pm\infty)=
  \Psi (k_1,
  f_1e^{-i\varepsilon_{k_1}t},\dots ,k_n,f_ne^{-i\varepsilon_{k_n}t}|\pm\infty)
 \end{equation}
 describes the evolution of a state corresponding to a collection of $n$ particles with wave functions
$f_1\phi_1e^{-i\varepsilon_{k_1}t}$,\dots ,$f_n\phi_n e^{-i\varepsilon_{k_n}t}$ as $t\to \pm\infty.$

 In what follows we use the notations $a_k(f)=\int f(\bp)a_k(\bp)d\bp$, $ a_k^+(f)=(a_k(\bar f))^*= \int  f(\bp)a_k^+(\bp)d\bp.$

 \subsection{The existence of in- and out-states from strong asymptotic commutativity}
 In this subsection we derive the existence of limit in~\eqref{eq:50.9} from strong asymptotic commutativity of the algebra $\hat {\cal {A}}(\omega).$  ( We assume that $B_i$ are good operators and that the $NO$ condition  is satisfied.) More precisely,   we assume that 
 for every  $n$ we have
 \begin {equation}\label {SAC}|| [\hat A_1(\bx,t), \hat A_2]||\leq \frac {C_n(t)}{1+||\bx||^n}\end {equation}
 where $A_1, A_2\in \cal A$ and
 $C_n(t)$ is a function of at most polynomial growth. In present section we are dealing with the case when this condition is satisfied for all $n$, however, one can weaken the condition( \ref {SAC}) assuming that it holds only for some $n>1$ (see the first version of this paper or  \cite {MO} for the proof).

 For simplicity we derive the existence of  the limit    in~\eqref{eq:50.9} in the the case when $t_1=...=t_n=t$. We introduce the notation $\Psi (t)$ for the vector
 \begin{equation}
  \label{VV}
  \Psi(k_1,f_1,\dots ,k_n,f_n|t,\dots ,t)=\hat B_{k_1}(f_1,t) \cdots \hat B_{k_n}(f_n,t)\Phi
 \end{equation}
 
 Then  the existence of the limit in  ~\eqref{eq:50.9} follows from the estimate 
 $\int dt||\dot \Psi(t)||<\infty.$
 
 The estimate we need  can be derived from
 the following fact:
 \begin {equation} \label{eq:50.17}
 \int \|[ \hat B_{k_i}(f_i,t), \dot{\hat B}_{k_j}(f_j,t)]\|dt<\infty.
 \end {equation}
 
 To check this we notice that $\dot \Psi(t)$ is a sum of $n$ terms; every term contains a product of several operators $\hat B$ and one operator $\dot {\hat B}.$ The estimate for the norm of $\dot \Psi(t)$ follows from~\eqref{eq:50.17},  (\ref {ON}) and the remark that the norms of operators $\hat B_{k_i}(f_i,t)$ are bounded.
We change the order of operators in the summand under consideration in such a way that $\dot {\hat B}_k$ is from the right.
 Then we notice that $\dot{\hat B}_k(f,t)\Phi=0$  as follows from (\ref {ON}).)

 The estimate ~\eqref{eq:50.17}  can be derived from the  following
 \begin {lemma}\label {L}
  Let us assume that   $\rm {supp}(\phi)$ is a compact set. Then  for large $|\tau|$ we have
$$ |(T_{\tau}\phi)({\bx})|<  C_n (1+|{\bx}|^{2}+\tau^2)^{-n}$$
where $ \frac {{\bx}}{\tau}\notin U_{\phi}$, the initial data $\phi=\phi ({\bx})$ is the Fourier transform of $\phi ({\bk})$, 
and $n$ is an arbitrary integer.

Here  $\rm {supp}(\phi)$ is the closure of the set of points were $\phi (\bk)\neq 0$, $U_{\phi}$ is a set of all points of the form $\nabla \epsilon (\bk)$
where $\bk$ belongs to a neighborhood of  $\rm {supp}(\phi)$,
$(T_{\tau}\phi)({\bx})= \int d\bk e^{i\bk\bx-i\epsilon (\bk)\tau}\phi(\bk)$,
 the function $\epsilon(\bk)$ is smooth.

 \end {lemma}
 
 To prove this lemma we can use the fact that the gradient of 
 $\bk\bx-i\epsilon (\bk)\tau$  with respect to $\bk$ does not vanish at the points we are interested in, hence this expression can be linearized in a small neighborhood of every such point by means of  an appropriate change of variables. Now we can  use a partition of unity on the support of $\phi (\bk)$ and apply Riemann-Lebesgue lemma.
 
 Another proof that can be used also to analyze the case when the support of the function  $\phi (\bk)$  is not compact is based on the following  
 formula
  
    \begin{equation}
        \int \exp(-i \sigma(\bp)) \chi(\bp) d\bp = \int \exp(-i\sigma(\bp))(L^n \chi)(\bp) d\bp, \label{37.1}
    \end{equation}
 $(L \chi)(\bp) = - div ({\bf u}(\bp) \chi(\bp))$ and $ {\bf u}(\bp) = \frac{\nabla \sigma(\bp)}{|\nabla \sigma(\bp)|^2}$ (see, for example,\cite {52}). We assume that     $\chi(\bp)$ is a smooth function with compact support, but it is sufficient to assume that it is a smooth fast decreasing function. 

    In the case at hand, we have
    $$\sigma(\bp) = \epsilon(\bp) t - \bp \bx; \quad \chi(\bp) = f(\bp);$$
    $${\bf  u}(\bp) = \frac{\bv(\bp)t - x}{|\bv(\bp)t - \bx|^2} = \frac 1 t \cdot \frac{\bv(\bp) - \frac \bx t}{|\bv(\bp) - \frac \bx t|^2}.$$
    If $\frac \bx t \not \in U_{f}$, it is easy to check that
    $$\sup_{\bp \in {\rm  supp} {\chi}} |D^{(\alpha)} {\bf u}(\bp) | \leq \frac C{|t|}$$
    (here $D^{(\alpha)} = \frac{\partial^{|\alpha|}}{\partial p_1^{\alpha_1} p_2^{\alpha_2} p_3^{\alpha_3}}, |\alpha| = \alpha_1 + \alpha_2 + \alpha_3$ and $C$, here and in the following, denotes a quantity independent of $\bx$ and $t$, though it may depend on other parameters, for example here it depends on $\alpha$ ).
    The inequality
    \begin{equation}
        \sup_{\bp \in {\rm supp} {\chi}} |(L^n \chi)(\bp)| \leq C |t|^{-n}. \label{37.2}
    \end{equation}
    follows.
    The formula~\eqref{37.1} and the inequality~\eqref{37.2} lead to the inequality $| \tilde f(\bx | t) | \leq \frac C {1+|t|^n}$, where $\frac{\bx}{t} \not \in U_f$ ($n$ is an arbitrary number).\footnote {In the proof, we have assumed that the function $\chi (p)$ is smooth. One can relax this condition assuming that this function is $m$ times continuously differentiable. Then we  have the inequality (\ref {37.2}) for $n=m.$}
 
  We assumed that  for good operator $B$ we have $\hat B\Phi=\Phi (\phi)$ where $\phi (\bp)$ does not vanish anywhere.  Notice that  the condition that $\phi (\bp)$ does 
 not vanish was not used in the proof of existence of the
limit  (\ref {eq:50.9}).  If this condition is violated the formula (\ref {eq:50.11})
remains correct, but it does not specify $in$- and $out$-operators  $a_{in}^+(g)$ and
$a_{out}^+(g)$ for  all $g$.  In what follows we  omit the condition that $\phi (\bp)$ does not vanish in the definition of good operator remembering that with modified definition
sometimes we need several good operators to calculate M\o ller matrices and scattering matrix.

\subsection {M\o ller matrices are isometric}
 The definition of $S_{\pm}$ that we gave specifies these operators as multi-valued maps (for example we can use different good operators in the construction and it is not clear whether we get the same answer).
 However, we can check that this map is isometric and every multi-valued isometric map is really single-valued. In particular, this means that the definition does not depend on the choice  of good operators. It follows also that the vector $\Psi (k_1, f_1, ..., k_n,f_n|\pm\infty)$ does not change when we permute the arguments $(k_i,f_i)$ and $(k_j,f_j).$

 To prove that the map is isometric we express the inner product of two vectors of the form $\Psi (t)$ in terms of truncated correlation functions.
 Only two-point truncated correlation functions survive in the limit $t\to \pm \infty.$
 This allows us to say that the map is isometric (see \cite{MO} for a more detailed proof in a slightly different situation).

\subsection {In- and out-operators}
 Let us define in- and out -operators by the formulas
 \[a_{in} (f)S_-=S_-a(f), a_{in}^+(f)S_-=S_-a^+(f),\]
 \[a_{out} (f)S_+=S_+a(f), a_{out}^+(f)S_+=S_+a^+(f).\]
 (For simplicity of notations we consider the case when we have only one type of particles.
 If we have several types of particles, in- and out-operators as well as the operators $a^+, a$ are labelled by a pair $(k,f)$ where $f$ is a test function and $k$ characterizes the type of particle.)
 These operators are defined on the image of $S_-$ and $S_+$ correspondingly.
 One can check that
 \begin{equation}\label{eq:50.11}
  a^+_{in}( f\phi)= \lim_{t\to-\infty} \hat B(f,t),\\
  a^+_{out}( f\phi)= \lim_{t\to\infty} \hat B(f,t),\\
 \end{equation}
 The limit is understood as a strong limit.
 It exists on the set of vectors of the form~\eqref{eq:50.10} (with $NO$ assumption for $d<3$).
 The proof follows immediately from the fact that taking the limit $t_i\to\infty$ in the vectors~\eqref{eq:50.9} we can first take the limit for $i>1$ and then the limit $t_1\to \infty.$

 The formula~\eqref{eq:50.11} can be written in the following way:
 \[a^+_{in}( f\phi)\Psi (f_1,\dots ,f_n|-\infty)=\Psi (f,f_1,\dots ,f_n|-\infty),\]
 \[a^+_{out}( f \phi)\Psi (f_1,\dots ,f_n|\infty)=\Psi (f,f_1,\dots ,f_n|\infty).\]
 Similarly,
 \begin{equation} \label{eq:50.12}
  a_{in}( f )\Psi (\phi ^{-1}f,f_1,\dots ,f_{n}|-\infty)= \Psi (f_1,\dots ,f_n|-\infty),
 \end{equation}
 \begin{equation} \label{eq:50.13}
  a_{out}( f )\Psi (\phi ^{-1} f,f_1,\dots ,f_{n}|\infty)= \Psi (f_1,\dots ,f_n|\infty).
 \end{equation}

 \subsection {Scattering marix}
Let us define the scattering matrix by the formula
 \[S=S_+^*S_-.\]

 If the operators $S_+$ and $S_-$ are unitary, we say that the theory has a particle interpretation.
 In this case (and also in the more general case when the image of $S_-$ coincides with the image of $S_+$) the scattering matrix is a unitary operator in $\cH_{as}$.
 Its matrix elements in the basis $ |\bp_1,\dots , \bp_n \ket=\frac {1}{n!} a^+(\bp_1)\dots a^+(\bp_n)\Phi$ (scattering amplitudes) can be expressed in terms of in- and out-operators:
 \begin{equation} \label{eq:50.14}
  S_{mn}(\bp_1,\dots ,\bp_m|\bq_1,\dots ,\bq_n)=\bra a^+_{in}(\bq_1)\dots a^+_{in}(\bq_n)\Phi,a^+_{out}(\bp_1) \dots a^+_{out}(\bp_m)\Phi\ket
 \end{equation}
In this formula and in what follows we  omit numerical factors $(m!)^{-1} (n!)^{-1}.$

 Effective cross-sections can be expressed in terms of the squares of scattering amplitudes.
 
 We will study  the expression  (\ref {eq:50.14}) in this section. ( Notice that we do not use the fact that the theory has particle interpretation in our considerations.)

 Only when $\omega$ is a ground state can one hope that the particle interpretations exists.
 In other cases instead of a scattering matrix and cross-sections one should consider inclusive scattering matrix and inclusive cross-sections (see Section 6).

 The formula~\eqref{eq:50.14} is proved only for the case when all values of momenta $\bp_i,\bq_j$ are distinct .
 (More precisely,  we should assume that  all vectors  $\bv (\bp_i)=\nabla \varepsilon (\bp_i), \bv (\bq_j)=\nabla \varepsilon (\bq_j)$ are distinct, but in the case when the function $\varepsilon (\bp)$  is strongly convex  it  is sufficient to assume that  $ \bp_i,\bq_j$ are distinct.) The formula ~\eqref{eq:50.14}   should be understood in the sense of generalized functions and as test functions we should take collections of functions $f_i(\bp_i),g_j(\bq_j)$ with non-overlapping $\overline{U(f_i)}, \overline {U(g_j)}$.
 Let us write this in more detail:
 \[S_{mn}(f_1,\dots , f_m|g_1,\dots ,g_n)=$$ $$ \int d^m\bp d^n\bq \prod f_i(\bp_i)\prod g_j(\bq_j) S_{mn}(\bp_1,\dots ,\bp_m|\bq_1,\dots ,\bq_n)=\]
 \[\bra a^+_{in}( g_1)\dots a^+_{in}( g_n)\Phi, a^+_{out}(\bar f_1)\dots a^+_{out}(\bar f_m)\Phi \ket\]
 Using~\eqref{eq:50.11}, we obtain
 \[S_{mn}(f_1,, f_m|g_1,\dots ,g_n)=\]
 \[\lim_ {t\to \infty,\tau\to -\infty} \bra \Phi|\hat B(\bar  f_m  \phi ^{-1},t)^* \dots \hat B( \bar f_1 \phi^{-1},t)^* \hat B( g_1 \phi^{-1},\tau) \dots \hat B(g_n\phi^{-1},\tau))\\|\Phi\ket\]
 \[\lim_ {t \to \infty, \tau \to -\infty} \omega(B(\bar f_m\bar  \phi ^{-1},t)^* \dots B(\bar  f_1 \bar \phi^{-1},t)^* B(g_1 \phi^{-1},\tau) \dots B( g_n\phi^{-1},\tau))\]
 where $B( f,t)^*=\int d\bx B^*(\bx,t)\overline {\tilde f(\bx,t)}.$

 Notice that in the same way we can obtain a more general formula
 \begin{equation}\label{eq:50.15}\begin {split}
  S_{mn}(f_1,\dots , f_m|g_1,\dots ,g_n)=                                                                                                                                                                    
  \lim_ {t_i\to \infty,\tau_j\to -\infty}\\ \omega(B_m( \bar f_m \phi_m ^{-1},t_m)^* \dots B_1(\bar  f_1 \phi_1^{-1},t_1)^* B_{m+1}(g_1 \phi_{m+1}^{-1},\tau_1)\dots B_{m+n}( g_n \phi_{m+n}^{-1},\tau_n))                            \\
 \end{split}\end{equation}
 where $B_i$ are different good operators and $B_i\Phi=\Phi(\phi_i)$. Due to $NO$ condition the ordering  of factors with $t_i$ tending to infinity is irrelevant. The same is true for factors with $\tau_j\to -\infty.$ This means that we can assume that the factors in the last line of (\ref {eq:50.15})  are time ordered.
 
 Using (\ref {GOO}) one can represent this formula in the  form 
 \begin {equation} \label {SOS} \begin{split}  S_{mn}(f_1,\dots , f_m|g_1,\dots ,g_n)=\\
\int d^{m+n}\bp 
 \lim _{t_i\to\infty,\tau_j \to -\infty}\bra \Phi|  f_m \bar \phi_m^{-1} e^{i\varepsilon_m(\bp_m)t_m}\hat  B_m (\bp_m,t_m)^* ...
   f_1 \bar\phi_1^{-1} e^{i\varepsilon_1(\bp_1)t_1} \hat  B_1 (\bp_1,t_1)^*\\ g_1 \phi_{m+1}^{-1}e^{-i\varepsilon _{m+1}(\bp_{m+1})\tau_1} \hat B_{m+1}(\bp_{m+1},\tau_n ) ... \phi_{m+n}^{-1}e^{-i\varepsilon _{m+n}(\bp_{m+n})\tau_n} \hat B_{m+n}(\bp_{m+n},\tau_n)|\Phi\ket\\ 
\end {split}\end{equation}
Equivalently we can write
\begin {equation} \label {SPS} \begin{split}  S_{mn}(\bp_1,\dots , \bp_m| \bp_{m+1}\dots ,\bp_{m+n})=
 \lim _{t_1,...,t_m\to\infty,t_{m+1},...,t_{m+n} \to -\infty}\\ \bra \Phi|  \bar \phi_m^{-1} e^{i\varepsilon_m(\bp_m)t_m}\hat  B_m (\bp_m,t_m)^* ...
  \bar \phi_1^{-1} e^{i\varepsilon_1(\bp_1)t_1} \hat  B_1 (\bp_1,t_1)^* \\\phi_{m+1}^{-1}e^{-i\varepsilon _{m+1}(\bp_{m+1})t_{m+1}}  \hat B_{m+1}(\bp_{m+1},t_{m+1} )...\\  \phi_{m+n}^{-1}e^{-i\varepsilon _{m+n}(\bp_{m+n}), t_{m+n}} \hat B_{m+n}(\bp_{m+n},t_{m+n})|\Phi \ket\\ 
   \end {split}\end{equation}
   
   Notice that in (\ref {SPS} ) we also can assume that the factors are time ordered. Later we will use this remark to express $S_{mn}$ in terms of Green functions  (LSZ formula).
   
   Taking in (\ref{eq:50.15}), (\ref {SOS} ) or in (\ref {SPS}) a limit with respect  to a part of variables $t_1,...,t_m$ and using (\ref {eq:50.11}) we obtain
    \begin{equation}\label{SSSS}\begin {split}
  S_{mn}(f_1,\dots , f_m|g_1,\dots ,g_n)=                                                                                                                                                                    
  \lim _{t_i\to\infty,\tau_j \to -\infty} \\ \bra \hat B_{m+1} ( g_1 \phi_{m+1}^{-1},\tau_1) \dots \hat B_{m+n}( g_n \phi_{m+n}^{-1},\tau_n))\Phi,\\  \hat B_1( f_1\phi_1^{-1},t_1)\dots B_r( f_r \phi_r ^{-1},t_r) a^+_{out}(f_{r+1}) ...a^+_{out}(f_m)\Phi \ket                          \\
 \end{split}\end{equation}
  \begin {equation} \label {SPSS} \begin{split}  S_{mn}(\bp_1,\dots , \bp_m| \bp_{m+1},\dots ,\bp_{m+n})=
 \lim _{t_1,...,t_m\to\infty,t_{m+1},...,t_{m+n} \to -\infty}\\ \bra \Phi|  a_{out}(\bp_m)...a_{out}(\bp_{r+1} )\bar\phi_r^{-1} e^{i\varepsilon_r(\bp_r)t_r}\hat  B_r (\bp_r,t_r)^* ...
  \bar \phi_1^{-1} e^{i\varepsilon_1(\bp_1)t_1} \hat  B_1 (\bp_1,t_1)^*\\  \phi_{m+1}^{-1}e^{-i\varepsilon _{m+1}(\bp_{m+1})t_{m+1}} \hat B_{m+1}(\bp_{m+1},t_{m+1} ) ... \phi_{m+n}^{-1}e^{-i\varepsilon _{m+n}(\bp_{m+n}), t_{m+n}} \hat B_{m+n}(\bp_{m+n},t_{m+n})|\Phi \ket\\ 
   \end {split}\end{equation}

\subsection {Scattering in theories with particle interpretation}

  In the proof of the existence of the limit  ~\eqref{eq:50.9} the assumption that $\hat B_k$ were good operators was used only to say that  $\dot{\hat B}_k(f,t)\Phi=0. $ If the theory has particle interpretation we can weaken this condition. 
  We assume that  the operators $B_k$ are smooth, the projection of $\hat B_k\Phi$ on the one-particle space has the form $\Phi_k (\phi_k)$ where $\phi_k$ is a non-vanishing function and the projection of  $\hat B_k\Phi$ on $\Phi$ vanishes.  Asymptotic commutativity allows us to move the factor with time derivative to the right; we prove that the expression we obtain is small for large $t$ (it is sufficient to prove that it is a summable function of $t$).
  As every vector in Hilbert space $\overline{\cal H}$ the vector $\hat B_k\Phi$ can be represented as
  $$\sum_{r\geq 0}\int d\bp_1...d\bp_r c_r(\bp_1,...,\bp_r)
  a_{in}^+(\bp_1)...a_{in}^+(\bp_r)\Phi.$$

  We  assumed that $\bra \Phi|\hat B_k|\Phi \ket =0$, hence
   we can represent ${\hat B}_k(f,t)\Phi$ in the form
  
  \begin {equation}\label {BP}
  \begin {split}{\hat B}_k(f,t)\Phi=\int d\bp\phi_k(\bp)f(\bp)+\\
  \sum_{r\geq 2} \int d\bp d\bp_1...d\bp_r e^ {-it(\varepsilon (\bp_1)+...\varepsilon (\bp_r)-\varepsilon (\bp))}c_r(\bp_1,...,\bp_r) f(\bp)a_{in}^+(\bp_1)...a_{in}^+(\bp_r)\Phi.\\
 \end{split}
 \end{equation}
  
  The first summand does not contribute to the time derivative of  ${\hat B}_k(f,t)\Phi$. The second summand contains a phase factor with large phase for $t\to -\infty$.  We assumed that operators $B_k$ are smooth; it follows that functions $c_r$ are smooth.
  Using this fact we can  prove that the contribution of the time derivative  of ${\hat B}_k(f,t)\Phi$ to $\dot\Psi(t)$ is small enough 
   in the limit $t\to-\infty.$  Then we can say that   limit  ~\eqref{eq:50.9}  exists and the expressions
  for matrix elements of scattering matrix (for example the formula (\ref {SPS}) ) remain correct if the operators $B_k$ are smooth operators such  that $\bra \Phi|\hat B_k|\Phi\ket=0$  and 
 the projection of $\hat B_k\Phi$ on  one-particle space  is  equal to $\Phi(\phi_k)$ where $\phi_k$ is a non-vanishing function.
    
  One can verify that the limit ~\eqref{eq:50.9}  does not depend  on the choice of operators $B_k$  also in the case when we do not assume that these operators are good. For good operators we   derived  this fact from the statement that the M\o ller matrices are isometric, the same proof works in  general case.  Again we express the inner product of two vectors of the form $\Psi(t)$ in terms of truncated correlation functions and notice that only two-point correlation functions of the form 
  \begin {equation}\label {PPP}\bra\Phi| {\hat B}_k^*(f,t){\hat B}_{k'}(f',t)|\Phi\ket=\bra {\hat B}_{k'}(f',t)|\Phi, {\hat B}_k(f,t)\Phi\ket\end{equation}
   survive in the limit $t\to \pm \infty.$ To finish the proof we  should calculate (\ref {PPP}) using (\ref {BP}) and apply  the fact that only the contribution of the first summand of (\ref {BP}) is relevant in the limit $t\to \pm \infty.$ 
  
  \subsection { Poincar\'e invariant theories}
  
  Let us consider theories where the translation symmetry can be extended to the symmetry with respect to Poincar\' e group (relativistic theories).  
  One can prove that in such theories scattering matrix is Lorentz-invariant; let us sketch the proof of this fact.

 The operator $\hat B(f,t)$ used in our construction can be written as an integral of
$\tilde f(\bx,t)B(\bx,t)$  where $\tilde f(\bx,t)$ is a positive frequency solution of the Klein-Gordon equation over the hyperplane $t= constant$. We will define the operator $\hat B(f,\rho)$  integrating the same integrand over another hyperplane $\rho.$  One can replace
 the operators $\hat B(f,t)$ by the operators $\hat B(f,\rho)$ in (\ref{eq:50.9}) and prove that the expression we obtained has a limit  if the hyperplanes $\rho_i$ tend to infinity in time direction
 (for example if they have form  $\alpha t+\ba \bx=constant $ and the constant  tends to infinity). This means that we can use $\hat B(f,\rho)$ in the definition of M\o ller matrices and scattering matrix.  The same arguments that were used to prove that the limit does not depend on the choice of good operators can be applied to verify that the new construction gives the same M\o ller matrices. This implies Lorentz invariance because the Lorentz group acts naturally on operators $\hat B(f,\rho).$
  
 For local relativistic theories with a mass gap one can derive asymptotic commutativity and cluster property from Haag-Araki  or Wightman axioms.   We obtain the existence and Lorentz-invariance of scattering   in such theories.
  
\section{Scattering matrix and inclusive scattering matrix in terms of Green functions}

 \subsection{Green functions and scattering. LSZ formula}
 Let us start with scattering of elementary excitations of ground state (of particles). In this case the scattering matrix can be expressed in terms of
 on- shell values of Green functions.
 The Green function in translation-invariant stationary state $\omega$ is defined by the formula
 $$G_n=\omega ( T(A_1(\bx_1, t_1) \dots A_r(\bx_r,t_r)))=$$
 $$\bra \Phi| T(\hat A_1(\bx_1, t_1) \dots \hat A_r(\bx_r,t_r))|\Phi \ket$$
 where $A_i\in \cal A$ and $T$ stands for time ordering. More precisely, this is a definition of Green function in $(\bx,t)$-representation, taking Fourier transform with respect to $\bx$ we obtain Green functions in $(\bp,t)$- representation, taking in these functions inverse Fourier transform with respect to $t$ we obtain Green functions in in $(\bp,\epsilon)$ -representation.
 Due to translation-invariance of $\omega$ we obtain that in $(\bx,t)$-representation the Green function depends on differences $\bx_i-\bx_j, t_i-t_j$, in $(\bp,t)$-representation it contains a factor $\delta(\bp_1 + \dots + \bp_r)$. Similarly in $(\bp, \epsilon)$ we have the same factor and the factor $\delta(\epsilon_1+ \dots +\epsilon_r).$
 We omit both factors talking about poles of Green functions.

 For $n=2$ in $(\bp,\epsilon)$-representation
$G_2$ has the form $$G(\bp_1,\epsilon_1|A,A')\delta(\bp_1+\bp_2)\delta (\epsilon_1+\epsilon_2).$$ Poles of function $G(\bp,\epsilon|A,A')$ correspond to particles, the dependence of the position of the pole on $\bp$ specifies the dispersion law $\varepsilon (\bp)$ (we consider poles with respect to the variable $\epsilon$ for fixed $\bp).$ This follows from K$\rm{\ddot a}$ll\'en-Lehmann representation, but this can be obtained also from the considerations below.

It follows from ( \ref {SPS}) that  to find scattering amplitudes we should consider asymptotic behavior of Green functions in $\bp,t$ representation as $t\to \pm \infty.$
This follows immediately from the remark that we can assume that the factors in (\ref {SPS}) are time ordered.  The formula (\ref {SPS}) implies this in the case when the Green function is constructed by means of good operators, but  it follows from the considerations in the Section 4.6 that this is true also in more  general case. The   asymptotic behavior of Green function in  $(\bp,t)$ representation is governed by the poles of Green function in $(\bp,\epsilon)$-representation  and by residues at these poles (by on-shell values  of Green function).  (Let us suppose that a function $\rho (t)$ has asymptotic
behavior $e^{-itE_{\pm}}A_{\pm}$ as $t\to \pm\infty$ (in other words there exist finite  limits $\lim _{t\to \pm \infty}e^{itE_{\pm}}\rho (t)=A_{\pm}$). Then the (inverse) Fourier transform  $\rho (\epsilon)$ has poles at the points $E_{\pm}\pm i0$ with residues 
$\mp 2\pi iA_{\pm}.$) 
 
 Hence the scattering amplitudes can be expressed in terms of on -shell values of Green functions (LSZ formula).
 
 Let us formulate this statement more precisely and give more detailed proofs.
 
 To simplify notations we consider the case when we have only one single-particle state $\Phi (\bp)$ with dispersion law $\varepsilon (\bp)$.
 We assume that the elements $A_i\in \cal A$ are chosen in such a way that the projection of $\hat A_i\Phi$ on $\Phi$ vanishes and the projection of this vector  on the one-particle space has the form $\Phi (\phi_i)=\int \phi_i(\bp)\Phi(\bp)d\bp$ where $\phi_i(\bp)$ is a non-vanishing function. Notice that for the case when $A_i$ is a good operator this notation agrees with the notation we used earlier.
 
 We introduce the notation $\Lambda_i(\bp)=\phi _i(\bp)^{-1}.$

 Let us consider Green function
 $$ G_{mn}=\omega(T(A_1^*(\bx_1, t_1)\dots A_m^*(\bx_m, t_m)A_{m+1}(\bx_{m+1}, t_{m+1})\dots A_{m+n}(\bx_{m+n}, t_{m+n}))$$
 in $(\bp,\epsilon)$-representation. It is convenient to change slightly the definition of  $(\bp,\epsilon)$-representation changing the signs of variables $\bp_i$ and $\epsilon_i$ for $1\leq i\leq m$ (for variables corresponding to the operators $A^*_i$).
 Multiplying the Green function in $(\bp,\epsilon)$-representation by
 $$ \prod_{1\leq i\leq m} \overline {\Lambda_i(\bp_i)}(\epsilon_i+\varepsilon (\bp_i))
  \prod _{m<j\leq m+n} \Lambda_j(\bp_j)(\epsilon_j-\varepsilon(\bp_j)).$$
 and taking the limit $\epsilon_i\to -\varepsilon(\bp_i)$ for $1\leq i\leq m$ and the limit
$\epsilon_j\to \varepsilon(\bp_j)$ for $m<j\leq m+n$ we obtain normalized on-shell Green function denoted by $\sigma_{mn}$. Deleting in the definition  of normalized on-shell Green function factors $\Lambda_j, \overline {\Lambda_i}$ we  obtain the definition of non-normalized on-shell Green function.

 We prove that {\it the normalized on-shell Green function  coincides with scattering amplitudes(LSZ formula): }
 \begin{equation}\label{eq:50.201}
  \sigma_{m,n}(\bp_1,\dots ,\bp_{m+n})=S_{mn}(\bp_1,\dots ,\bp_m| \bp_{m+1},\dots \bp_{m+n}).
 \end{equation}
 First of all we notice that the on-shell Green function can be expressed in terms of the asymptotic behavior of the Green function in $(\bp, t)$-representation: if for $t\to\pm \infty$ and fixed $\bp$ this behavior is described by linear combination of exponent $e^{i\lambda t}$ then the location of poles is determined by the exponent indicators $\lambda$ and the coefficients in front of exponents determine the residues. Using this statement one can  show that

 \begin{equation}
  \lim_ {t_i\to \infty,\tau_j\to -\infty} \omega(B_m( \bar f_m\bar  \phi_m ^{-1},t_m)^* \dots A_1(\bar  f_1\bar \phi_1^{-1},t_1)^* A_{m+1}(g_1 \phi_{m+1}^{-1},\tau_1)\dots A_{m+n}( g_n \phi_{m+n}^{-1},\tau_n))                            \\
 \end{equation} where $A_i$ are good operators can be expressed in terms of normalized on-shell Green function as
 $$\int  d^{m+n}\bp  f_1(\bp_1)...f_m(\bp_m)g_1(-\bp_{m+1})...g_n(-\bp_{m+n})
  \sigma_{m,n}(\bp_1,\dots ,\bp_{m+n}).$$  Using~\eqref{eq:50.15} we obtain~\eqref{eq:50.201} in the case when $A_i$ are good operators. 
  Equivalent  (and more transparent ) way to prove this statement is to use (\ref {SPS}).
  
  If we assume that operators $\hat A_i$ are smooth and the theory has particle interpretation we can  use the results of Section 4.6. It follows from these results that (\ref {SPS}) and other  formulas of Section  4.5  can be applied also in the case when these operators  are not good. This means that the  LSZ formula (\ref {eq:50.201}) remains correct in this case.

 \subsection{Generalized Green functions. Inclusive scattering matrix}
 Let us define generalized Green functions (GGreen functions) in the state $\omega$ by the following formula where $B_i\in \cal A$:

 $$G_n=\omega (MN)$$
 where $$N= T( B_1(\bx_1, t_1)\dots B_n(\bx_n,t_n))$$ stands for chronological product (times decreasing)
 and
 $$M=T^{opp}(B^*_1(\bx'_1,t'_1)\dots B_n^*(\bx'_n,t'_n))$$ stands for antichronological product (times increasing). Notice that taking the Hermitian conjugate of chronological product we obtain antichronological product, hence
 $$ M=(T(B_1(\bx'_1,t'_1)\dots B_n(\bx'_n,t'_n))^*.$$

 One can give another definition of GGreen functions introducing the operator
 $$ Q=T(B_1(\bx_1,t_1)\dots B_n(\bx_n, t_n)\tilde B_1(\bx'_1,t'_1)\dots \tilde B_n(\bx'_n,t'_n)).$$ where the operators $B_i,\tilde B_i$ act on the space of linear functionals on $\cal A.$
 (Recall~\eqref{eq:50.61},~\eqref{eq:50.611} that operators $B$ and $\tilde B$ act on linear functionals defined on $\cal A$; they transform $\omega (A)$ into $\omega(AB)$ and in $\omega (B^*A)$ correspondingly.)
 It is easy to check that
 $$G_n=(Q\omega)(1)$$
or, in bra-ket notations,
$$\langle 1| T(B_1(\bx_1,t_1)\dots B_n(\bx_n, t_n)\tilde B_1(\bx'_1,t'_1)\dots \tilde B_n(\bx'_n,t'_n))|\omega\rangle.$$
 Let us define inclusive $S$-matrix  as  on-shell GGreen function. We will show that  in the case when the theory has particle interpretation inclusive cross-section can be expressed
 in terms of inclusive $S$-matrix.

 The inclusive cross-section  of the process $(M,N)\to (Q_1 ...,Q_m)$ is defined as a sum (more precisely a sum of integrals)  of effective cross-sections  of the processes $(M,N)\to (Q_1,...,Q_m, R_1, ..., R_n$ over all possible $R_1,...,R_n.$ If the theory does not have particle interpretation this  formal definition of inclusive cross-section does not work, but still the inclusive cross-section can be defined in terms of  probability of the process
$(M,N\to (Q_1,...,Q_n+$ something else) and expressed in terms of inclusive $S$-matrix.

Let us consider the expectation value
\begin{equation}
 \label{eq:NU}
 \nu (a^+_{out, k_1}(\bp_1)a_{out, k_1}(\bp_1)\dots a^+_{out,k_m}(\bp_m)a_{out,k_m}(\bp_m))
\end{equation}
where $\nu $ is an arbitrary state.
This quantity is the probability density in momentum space for finding $m$ outgoing particles of the types $k_1,\dots ,k_n$ with momenta $\bp_1,\dots ,\bp_m$ plus other unspecified outgoing particles.
It gives inclusive cross-section if $\nu=\nu(t)$ describes the evolution of a state represented as a collection of incoming particles.

It will be convenient to consider more general expectation value
\begin{equation}
 \label{NUU}
 \nu (a^+_{out, k_1}(f_1)a_{out, k_1}(g_1)\dots a^+_{out,k_m}
 (f_m)a_{out,k_m}(g_m))
\end{equation}
where $f_i,g_i$ are test functions.
This expression can be understood as a generalized function
\[
 \nu (a^+_{out, k_1}(\bp_1)a_{out, k_1}(\bq_1)\dots a^+_{out,k_m}(\bp_m)a_{out,k_m}(\bq_m));\]
we get~\eqref{eq:NU} taking $\bp_i=\bq_i.$

As earlier we assume that we have several types of (quasi)particles $\Phi_k(\bp)$ obeying $\bP\Phi_k=\bp\Phi_k(\bp), H\Phi_k(\bp)=\varepsilon_k(\bp) \Phi_k(\bp)$ where the functions $\varepsilon_k(\bp)$ are smooth and strictly convex.
Good operators $B_k\in \cA$ obey $\hat B_k\Phi=\Phi (\phi_k).$ The operator $\hat B_k(f,t)$ where $f$ is a function of $\bp$ defined as $\int \tilde f(\bx,t)\hat B_k(\bx,t)d\bx$ ( as always $\tilde f(\bx,t)$ is a Fourier transform of
$f(\bp)e^{-i\varepsilon_k(\bp)t}$ with respect to $\bp).$

Now we can calculate (\ref {NUU}).
First of all we take as $\nu$ the state corresponding to the vector (\ref {eq:50.101}).
We are representing this vector in terms of good operators $B_i$
using the formula

$$\Psi(k_1,f_1,\dots k_n,f_n|\infty)= \lim _{t\to -\infty}\Psi(k_1,f_1,\dots ,k_n,f_n|t)=$$ $$\lim _{t\to -\infty}\hat B_{k_1}(f_1,t) \cdots \hat B_{k_n}(f_n,t)\Phi.$$

The corresponding state $\nu$ considered as as linear functional on $\cal A$ can be expressed in terms of the state $\omega$ corresponding to $\Phi.$ ( We should use the remark that the state corresponding to the vector $B\Phi$ can be written as $\tilde B B\omega.$) Expressing the out -operator by the formula(\ref {eq:50.11}) we obtain the expression of  (\ref {NUU}) in terms of GGreen functions on shell. 

Let us sketch another derivation  of the relation between the GGreen functions on shell and  inclusive cross sections. We start with representation of GGreen function
$G_n=\omega (MN)=\langle \Phi| MN |\Phi \rangle$ in the form 
$$\sum _l \int d\bp_1...d\bp_l \langle \Phi|M |\bp_1...\bp_l\rangle\langle \bp_1...\bp_l|N|\Phi\rangle$$ where $|\bp_1...\bp_l\rangle=\frac {1}{l!} |a^+_{out}(\bp_1)...a^+_{out}(\bp_l)\Phi\rangle$ is a generalized basis  of $\bar {\cal H}.$ ( Recall that we assume that the theory has particle interpretation.) On shell both $\langle \Phi|M |\bp_1...\bp_l\rangle$ and $\langle \bp_1...\bp_l|N| \Phi\rangle$ can be represented in terms of scattering matrix; this follows  from (\ref {SPSS}). Using these representations.  we obtain an expression of GGreen function on shell in terms of scattering matrix. This expression immediately leads to relation between GGreen functions on shell and inclusive cross-sections.

 {\bf Appendix}

We have emphasized that particles can be unstable and quasiparticles are almost always unstable. Nevertheless we disregarded instability in our considerations. This approach can be justified by the following remark: If a (quasi)particle has lifetime $T$ then for the times $t<<T$ it can be considered as a stable particle (hence in the case when the collision time is much less than $T$ we can consider scattering). We will give a heuristic proof of this statement.

Unstable (quasi)particles correspond to complex poles
$\varepsilon (\bp)+i\Gamma (\bp)$ of the two-point Green function $G(\bp,\epsilon|A,A').$ We know that  in the case of stable particles we can construct an operator $B$ transforming $\Phi$ into one-particle state ( a good operator) using the formula $B= \int \alpha (\bx,t) A(\bx,t) d\bx dt $ where $A\in \cA$ and the projection of $A\Phi$ onto one-particle state does not vanish.
 Namely we can assume that the function
$\hat\alpha(\bp,\omega)$ ( the Fourier transform of $\alpha$)  does not vanish only when $\omega$ is close  to $\varepsilon (\bp).$  (This is a particular case of the construction  we have used.) The same construction can be applied in the case of almost stable (quasi)particles ($T=\Gamma^{-1}>>0$). It seems that in  this case the operator $B$ is almost good in the following sense. Recall  that for a good operator $B(f,t)\Phi$ does not depend on $t$, hence the derivative with respect to $t$ vanishes. In our  case this derivative does not vanish, but it is small for $t<<T.$ This allows us to give an estimate for $\frac {d\Psi}{dt}$ where  $\Psi(t)$ is defined in Section 4.2. All above considerations can be repeated, but we cannot take the limit 
$t\to \infty$; for unstable particles we should always assume  that $t<<T$ .  We  obtain that the same picture as  for stable particles is approximately correct for unstable particles if $t<<T.$

{\bf Acknowledgements} I   would like to thank Simons Center, IFT (Sao Paulo) and IHES ( Bures-sur-Yvette) for their hospitality). I am indebted to N. Berkovits, D. Buchholz, T. Damur, A. Kamenev, Z. Komargorodsky, M. Kontsevich,  G.Lechner, A. Mikhailov, N. Nekrasov,  M. Rangamani for useful discussions.

 
\end {document}